\documentclass[prd, twocolumn, nofootinbib, floatfix]{revtex4} 

\usepackage{amsmath}
\usepackage{graphicx}
\usepackage{dcolumn}
\usepackage{bm}
\usepackage{epsfig}
\usepackage{amssymb,latexsym,mathrsfs}
\usepackage{graphicx}
\usepackage{color}
\usepackage{hyperref}
\usepackage[sort&compress]{natbib}
\usepackage{xfrac}

\hypersetup{
    colorlinks=true,
    linkcolor=red,
    citecolor=blue,
} 

\newcommand{\be}{\begin{equation}}
\newcommand{\ee}{\end{equation}}
\newcommand{\bs}{\begin{split}} 
\newcommand{\bea}{\begin{eqnarray}}
\newcommand{\eea}{\end{eqnarray}}

\hypersetup{
    bookmarks=true,         
    unicode=false,          
    pdftoolbar=true,        
    pdfmenubar=true,        
    pdffitwindow=false,     
    pdfstartview={FitH},    
    pdftitle={My title},    
    pdfauthor={Author},     
    pdfsubject={Subject},   
    pdfcreator={Creator},   
    pdfproducer={Producer}, 
    pdfkeywords={keyword1} {key2} {key3}, 
    pdfnewwindow=true,      
    colorlinks=false,       
    linkcolor=red,          
    citecolor=green,        
    filecolor=magenta,      
    urlcolor=cyan           
}

\newenvironment{itemize*}
  {\begin{itemize}
    \setlength{\itemsep}{0pt}
    \setlength{\parskip}{0pt}}
  {\end{itemize}}

\newenvironment{enumerate*}
  {\begin{enumerate}
    \setlength{\itemsep}{0pt}
    \setlength{\parskip}{0pt}}
  {\end{enumerate}}

\newenvironment{description*}
  {\begin{description}
    \setlength{\itemsep}{0pt}
    \setlength{\parskip}{0pt}}
  {\end{description}}

\def\ben{\begin{enumerate*}}
\def\een{\end{enumerate*}}
\def\bi{\begin{itemize*}}
\def\ei{\end{itemize*}}
\def\bd{\begin{description*}}
\def\ed{\end{description*}}
\def\be{\begin{equation}}
\def\ee{\end{equation}}
\def\bea{\begin{eqnarray}}
\def\eea{\end{eqnarray}}
\def\bfl{\begin{flushleft}}
\def\efl{\end{flushleft}}

\newcommand{\gev}{~\mbox{GeV}}
\newcommand{\mev}{~\mbox{MeV}}
\newcommand{\tev}{~\mbox{TeV}}

\newcommand{\rhos}{\rho_\sigma}
\newcommand{\rhor}{\rho_R}

\newcommand{\rhom}{\rho_X}

\newcommand{\gam}{\Gamma_\sigma}
\newcommand{\sgm}{\sigma}
\newcommand{\ds}{\delta_\sigma}

\newcommand{\dx}{\delta_X}
\newcommand{\ts}{\theta_\sigma}

\newcommand{\tx}{\theta_X}

\newcommand{\ra}{\rho_{(\alpha)}}
\newcommand{\wa}{w_{(\alpha)}}
\newcommand{\ca}{c_{(\alpha)}}

\newcommand{\da}{\delta_{(\alpha)}}
\newcommand{\ta}{\theta_{(\alpha)}}

\def\nn{\nonumber}
\def\bec{\begin{center}}
\def\eec{\end{center}}
\def\beq{\begin{eqnarray}}
\def\eeq{\end{eqnarray}}
\def\fr{\frac}

\begin{document}

\title{Non-thermal WIMPs and Primordial Black Holes}

\author{Julian Georg}
\email{jsgeorg@syr.edu}
\author{Gizem \c{S}eng\"or}
\email{gsengor@syr.edu}
\author{Scott Watson}
\email{gswatson@syr.edu} 
\affiliation{Department of Physics, Syracuse University, Syracuse, NY 13244, USA}

\date{\today}

\begin{abstract}
Non-thermal histories for the early universe have received notable attention as they are a rich source of phenomenology, while also being well motivated by top-down approaches to beyond the Standard Model physics. 
The early (pre-BBN) matter phase in these models leads to enhanced growth of density perturbations on sub-Hubble scales.
Here we consider whether primordial black hole formation associated with the enhanced growth is in conflict with existing observations.  
Such constraints depend on the tilt of the primordial power spectrum, and we find that non-thermal histories  are tightly constrained in the case of a significantly blue spectrum.  Alternatively, if dark matter is taken to be of non-thermal origin we can restrict the primordial power spectrum on scales inaccessible to CMB and LSS observations. We establish constraints for a wide range of scalar masses (reheat temperatures) with the most stringent bounds resulting from the formation of $10^{15}$ g black holes.  These black holes would be evaporating today and are constrained by FERMI observations.
We also consider whether the breakdown of the coherence of the scalar oscillations on sub-horizon scales can lead to a Jean's pressure 
preventing black hole formation and relaxing our constraints. Our main conclusion is that primordial black hole constraints, combined 
with existing constraints on non-thermal WIMPs, favor a primordial spectrum closer to scale invariance or a red tilted spectrum. 
\end{abstract}
\maketitle
\thispagestyle{empty}

\section{Introduction}
The post-inflationary universe prior to Big Bang Nucleosynthesis (BBN) lacks any direct connection to  cosmological observations. Given the absence of data, it is typically assumed that inflationary reheating quickly led to a thermal universe prior to BBN.  Then one calculates the consequences -- such as the relic abundance of thermal dark matter (DM) with Weakly Interacting Massive Particles (WIMPs) being a feasible candidate.

Non-thermal histories provide a well-motivated alternative \cite{Acharya:2008bk,Acharya:2009zt,Kane:2015jia}.
These theories arise naturally in supergravity and string theory based approaches 
to beyond the Standard Model physics.  
Non-thermal histories lead to a universe that is matter dominated prior to BBN due to the coherent 
oscillations of scalar (moduli) fields, which decay late due to their weak (gravitational strength) couplings.  
They also provide a rich and interesting source of DM phenomenology.
In these models thermal WIMPs freeze-out during the scalar dominated period. Later, when the scalar fields decay and reheat the universe again, they 
provide an additional source of DM in the form of non-thermal WIMPs \cite{Kane:2015jia}.  The non-thermal WIMPs typically
provide the dominant source of DM and their self annihilation rates can be many orders of magnitude larger than that of thermal WIMPs.  This leads to new expectations for direct and indirect detection experiments as well as other probes of the microscopic nature of DM \cite{Kane:2015jia}.  In addition, contrary to a thermal history, DM and scalar perturbations
will grow during the non-thermal phase \cite{Erickcek:2011us,Fan:2014zua}.  This can lead to an additional enhancement of DM sub-structure on small scales and has important implications for indirect detection signals and the process of structure formation \cite{Erickcek:2011us,Fan:2014zua,Erickcek:2015jza,Erickcek:2015bda,Kane:2015qea}.

In this paper, we want to consider whether the growth of scalar perturbations during the non-thermal phase leads to the over-production of Primordial 
Black Holes (PBHs).  Observational constraints on PBHs and their evaporation products provide a way in which to probe the nature of the post inflationary universe prior to BBN\footnote{The literature 
on PBHs is vast, we refer the reader to \cite{Khlopov:2008qy,Carr:2009jm,Green:2014faa} for reviews.}.
The formation rate of PBHs depends on the evolution of the universe and the primordial power spectrum on scales that are not probed by Cosmic Microwave Background (CMB) and Large Scale Structure (LSS) observations.
Observations favor a primordial `red spectrum' with increasing power on large scales and with a scalar tilt $n=0.968 \pm 0.006$ \cite{Ade:2015lrj}.  
For a strictly red spectrum, the average initial amplitude of perturbations at horizon entry will be too weak to collapse to form PBH \cite{Carr:1994ar}.
However, CMB and LSS observations only probe the 
inflationary potential around  $50$ to $60$ e-folds before the end of inflation and within a window of about nine e-folds.
This corresponds to observable scales $k^{-1} \simeq 1$ -- $10^4$ Mpc, telling us nothing about the primordial spectrum on smaller scales.
In fact, axion, hybrid, and hill-top inflation models all provide examples of models that can agree with CMB scale observations and still predict more power on small scales -- a so-called `blue spectrum' $n>1$ toward the end of inflation.  
Given our lack of knowledge of the primordial spectrum and of the post-inflationary expansion, PBH formation is an important possibility and can provide meaningful constraints.

The enhanced growth of density perturbations during a non-thermal history, leads to the prediction that PBHs can form on sub-Hubble scales.  This is in stark contrast 
to the situation in a radiation dominated universe where density perturbations experience logarithmic growth and PBHs can only form immediately as modes enter the horizon -- on smaller scales 
the process is halted by the Jean's pressure resulting from radiation. In a matter dominated universe the Jean's pressure vanishes and PBH formation on small scales becomes possible \cite{Khlopov:1985jw,Polnarev:1986bi}.

In this paper we consider constraints on non-thermal histories arising from the formation of PBHs.
In the next section we review how the enhanced growth of perturbations arises from an expansion dominated by an oscillating scalar field.  In Section \ref{sec:constraints} we establish the expected mass fraction of PBHs 
resulting from the enhanced growth of perturbations on sub-Hubble scales 
and establish the expected mass range. We then consider the constraints on the primordial power spectrum that result by demanding a non-thermal phase prior to BBN.  
In Section \ref{secDM} we combine our PBH constraints with existing constraints on non-thermal WIMPs in SUSY based models.  In particular, we focus on the wino-like neutralino and find that combining existing indirect detection constraints with our PBH bounds imply the scalar tilt is bounded by $n<1.25$ at high reheat temperatures, but this constraint softens to about $n<1.4$ at lower reheat temperatures due to the duration of the matter phase.  The longer the phase, the weaker the constraints.  In Section \ref{breakdown} we question whether treating scalar oscillations as `dust' is always appropriate in these models. It is known that scalar coherence can break down on sub-Hubble scales resulting in a Jean's pressure that can disrupt PBH formation, and so weaken our constraints.  In the last section we conclude. 
\section{Growth of Perturbations in the Non-thermal Phase}

\subsection{Background Evolution}
We are interested in the background evolution following the end of inflation.
In non-thermal histories, once the scalar mass becomes comparable to the expansion rate
coherent oscillations of the scalar will lead to an effectively matter dominated phase.  Within this regime we can describe 
the cosmological background as a system of three interacting fluids with energy densities $\rho_\sigma$, $\rho_R$, and $\rho_X$ corresponding to the scalar, radiation, and DM (DM) sectors, respectively.
The relevant equations are then\footnote{We work with metric signature $-,+,+,+$ and with $\hbar=c=k=1$ and reduced Planck mass $m_p=2.44 \times 10^{18} \gev$. Dots denote derivatives with respect to cosmic time.} \cite{Giudice:2000ex,Erickcek:2011us,Fan:2014zua}
{\small
\bea
 \dot{\rho}_\sgm &=& -3H\rhos - \gam\rhos, \label{BGE1} \\
\dot{\rho}_ R&=& -4H\rhor +(1-B_X) \gam \rhos +\frac{\langle \sigma v \rangle}{m_X}\left[\rhom^2-\tilde{\rho}_{X}^2\right],\;\;\;\;\label{BGE2}  \\
\dot{\rho}_X &=&-3H\rhom +B_X \gam\rhos -\frac{\langle \sgm v \rangle}{m_X}\left[\rhom^2-\tilde{\rho}_{X}^2\right], \label{BGE3} 
\eea
{\normalsize where} $\gam \sim m_{\sigma}^3/m_p^2$ is the scalar decay rate and $B_X$ is the branching ratio for decay to DM and we assume all other decays result in relativistic particles.  The DM self annihilation rate is $\langle \nn\sgm v \rangle$, and we denote its energy density in equilibrium as $\tilde{\rho}_{X}$.

We will be interested in the regime with DM non-relativistic $T \ll m_X$ and so we can neglect the equilibrium terms in \eqref{BGE2} and \eqref{BGE3}, since they will be Boltzmann suppressed. The radiation density is given by the temperature as  $\rhor=  {\pi^2 g_*T^4}/{30}$ and we must take care to track the non-standard relation between the temperature and expansion rate resulting from entropy production during scalar decays \cite{Giudice:2000ex}.
The Friedmann equations are
\bea
3 H^2 m_p^2 &=& \sum_\alpha \rho_{(\alpha)}, \label{FRW0}\\
2\dot{H} m_p^2&=& -\sum_\alpha ( \rho_{(\alpha)}+p_{(\alpha)}) ,\label{FRW}
\eea
where $\alpha$ runs over the values $\alpha=\sigma,R,X$ for each fluid.  

In the presence of DM annihilations the background evolution was found in \cite{Fan:2014zua}, which was in excellent agreement with the earlier studies \cite{Giudice:2000ex,Erickcek:2011us} where DM annihilations were absent. 
Deep within the non-thermal phase the scalar energy density evolves as expected $\rho_\sigma \sim 1/a^3$, whereas radiation and DM matter scale as 
$\sim 1/a^{3/2}$ due to the changing entropy as a result of scalar decays. The behavior of the system can be seen in Figure \ref{fig1}, where we plot the relative contribution of each fluid to the total energy density and we also include the scaling of the radiation temperature as compared with the DM mass (recall $T \ll m_X$).  Given the background evolution we now consider the growth of cosmological perturbations.

\begin{figure}[t!]
\begin{center}
\includegraphics[scale=0.43]{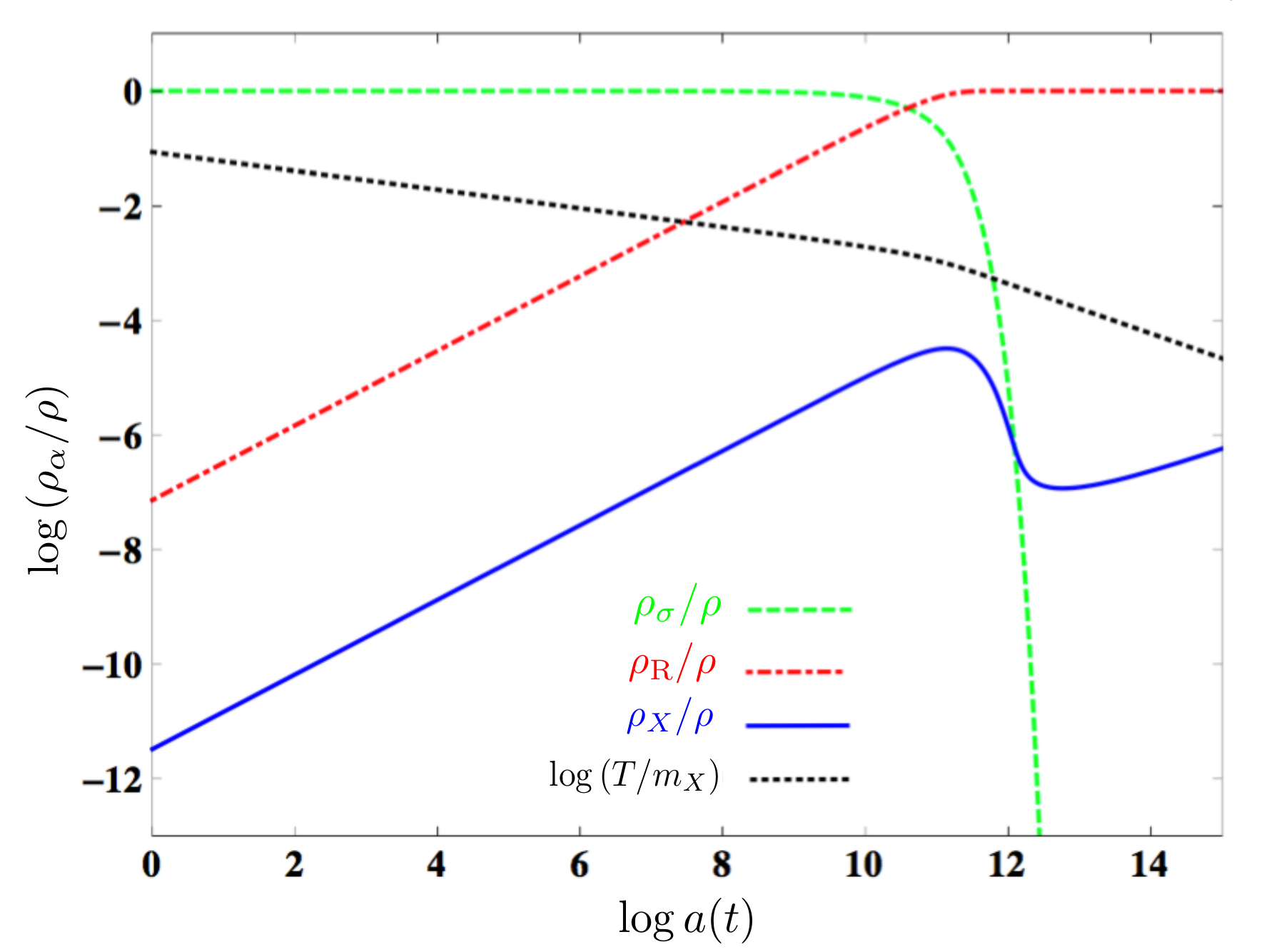}
\end{center}
\caption{\label{fig1} Evolution of the background energy densities compared to the total density $\rho$ for a non-thermal history. 
In the figure we take $m_X=500 \gev, B_X=1/3$,  $g_*=30$, and $\langle \sigma v \rangle=3\times 10^{-8} \gev^{-2}$. We also chose the initial dimensionless decay rate as $\gam/H_0 \simeq 2\times 10^{-7}$ and the scalar mass $m_\sigma=10^6 \gev$. The universe becomes radiation dominated at $\log a \simeq 11$ and at a temperature of around $700 \mev$. We refer the reader to \cite{Fan:2014zua} for more details. }
\end{figure}

\subsection{Growth of Perturbations \label{sectiongrow}}
The evolution of cosmological perturbations in a non-thermal history were studied in \cite{Fan:2014zua}.
In this paper we are interested in the growth of the scalar perturbations and whether this growth can lead to black hole formation.  Working in longitudinal gauge the scalar metric perturbation is given by
\be
ds^2=-\left(1+2 \Phi \right) dt^2 + a(t)^2\left(1-2 \Phi \right) d\vec{x}^2.
\ee
where we have assumed vanishing anisotropic stress.
As a consequence of the Einstein equations the {\em total} energy-momentum tensor is covariantly conserved
\begin{equation}
\nabla_{\mu}T^{\mu\nu}= \sum_{\alpha=\sigma, X,R} \nabla_{\mu}T^{\mu \nu}_{(\alpha)}=0,
\end{equation}
whereas each component $T^{\mu \nu}_{(\alpha)}$ may not be due to energy and particle transfer between sectors (e.g. from scalar decays and DM annihilations). 

Introducing fractional density perturbations $\delta_{(\alpha)} \equiv \delta \rho_{(\alpha)} / \rho_{(\alpha)}$ and defining the velocity perturbation for each fluid as $\theta_{(\alpha)}=a^{-1}\nabla^{2}v_{(\alpha)}$ we find a system of 
first order differential equations for each fluid
\bea
\nn \dot{\delta}_{(\alpha)}&+&3H(\ca^2-\wa)\da +(1+\wa)\left(\fr{\ta}{a}-3\dot{\Phi}\right) \\
\nn &=&-Q_{(\alpha)}\left(\ds - \da + \Phi\right) +P_{(\alpha)}\left(2\dx-\da +\Phi \right), \\
\nn \dot{\theta}_{(\alpha)}&+&H\ta -\frac{k^2\ca^2}{a (1+\wa)} \da -3H\wa\ta \\
&=& \fr{k^2}{a}\Phi  -Q_{(\alpha)} \Theta^{\sigma}_{(\alpha)}+ P_{(\alpha)}\Theta^{X}_{(\alpha)}, \label{fluid_eoms}
 \eea
where we again neglect DM equilibrium terms ($\tilde{\rho}_X=0$), we have used the background equations of motion, and we define
\bea
\nn Q_{(\alpha)}&\equiv&\gam ^{(\alpha)}\fr{\rhos}{\ra}, \\
\nn P_{(\alpha)}&\equiv&\fr{\langle \sgm v \rangle^{(\alpha)}}{m_X \ra}\rhom^2, \\
\nn \Theta_{(\alpha)}^\sigma &\equiv& \left[\fr{\ts}{1+\wa}-\ta\right], \\
\nn \Theta_{(\alpha)}^X &\equiv& \left[\fr{\tx}{1+\wa}-\ta\right],
\eea
where there is no sum over repeated indices. 
We are interested in the growth of scalar field perturbations, so assuming that the scalar background evolves as pressureless matter ($w_\sigma=0$) and that the perturbations have vanishing sound speed ($c_\sigma=0$) we find 
\bea \label{scalar1}
\dot{\delta}_\sigma&+&a^{-1}\theta_\sigma -3 \dot{\Phi} = - \Gamma_\sigma \Phi, \\
\dot{\theta}_\sigma&+&H \theta_\sigma=k^2a^{-1} \Phi. \label{scalar2}
\eea
The time-time component of the perturbed Einstein equation is
\bea
\label{EE1}
\left( \frac{k^2}{3 a^2 H} + H \right) \Phi + \dot{\Phi} &=&-\fr{1}{6H m_p^{2}}\sum_{\alpha} \rho_{(\alpha)} \delta_{(\alpha)}, \nn \\
& \simeq &-\fr{1}{2} H \delta_{(\sigma)},
\eea
where in the second step we use $\rho_\sigma \gg (\rho_X, \rho_R)$ and the background equation $3 H^2 m_p^2 \simeq \rho_\sigma$.  Given adiabatic initial conditions (isocurvature is negligible) the solution on sub-Hubble scales ($k \gg aH$) is $\delta_\sigma \sim a(t)$ and $\Phi \sim \Phi_0$ (constant) \cite{Fan:2014zua}.
That is, scalar field perturbations grow linearly during the non-thermal phase in contrast to the case in a 
thermal history where they only grow logarithmically\footnote{This is a familiar result from standard cosmology and explains why the majority of structure formation is expected to take place after the time of radiation / matter equality when the universe becomes matter dominated \cite{Mukhanov:2005sc}.}.

Given this enhanced growth of scalar perturbations on sub-Hubble scales we now examine whether this can result in an overproduction of primordial black holes.

\section{Primordial Black Hole Constraints \label{sec:constraints}}
Given the mass $M$ contained in a sphere of radius ${\lambda}/{2}=\pi a / k$ 
\be \label{Mdef}
M=\frac{4\pi}{3}\left(\frac{\lambda}{2}\right)^3\rho(t),
\ee
and assuming a primordial spectrum of nearly Gaussian fluctuations, the fraction
of the universe in PBHs at scale $M$ 
is expected to be\footnote{This calculation has been refined over the years, see e.g. \cite{Young:2014ana} and references within, but Carr's original result still provides an accurate enough estimate for many considerations.} \cite{Carr:1975qj}
\be  \label{rdubeta}
\beta_0(M) \simeq \delta_M(t_H) \exp\left( -\frac{w^2}{2 \delta^2_M(t_H)}\right),
\ee
where $w = p / \rho$ is the equation of state ($w>0$), 
$t_H$ is the time of Hubble radius crossing $k_H=a_H H $ and $\delta_M \equiv \delta M / M$ is the rms mass fluctuation.
This probability has an intuitive explanation.  As a mode enters the Hubble radius it will undergo growth due to the influence of gravity, whereas
this will be halted below the Jean's length due to the pressure $p=w\rho$.  Thus, a mode must enter with a large enough amplitude $\delta_M$ so 
that the initial perturbation can grow to $\delta_M(t_H)\sim{\cal O}(1)$ and break away from the background expansion before 
pressure effects prevent further collapse. 
For $w\sim{\cal O}(1)$ (as in a radiation dominated universe) the Jean's radius nearly coincides with the Hubble radius and so only perturbations with large initial amplitude and near the Hubble scale  will collapse to form PBHs.

\begin{table*}[t]
\caption{\\ Representative values for Non-thermal WIMP Models \label{table1}}
\vspace{-0.0in}
\be
\arraycolsep=1.4pt\def\arraystretch{1.7}
\begin{array}{|c|c|c|c|c|}\hline  & \;\;\; \mbox{Low reheat (MeV)} \;\;\;&\;\;\; \mbox{High reheat (TeV)} \;\;\;& \;\;\; \mbox{Low reheat (MeV)} \;\;\;&\;\;\; \mbox{High reheat (TeV)} \;\;\; \\\hline T_r & 3.3 \mev & 1.1 \tev & 3.3 \mev & 1.1 \tev \\\hline \Gamma & 4.8 \times 10^{-24} \gev & 2.4 \times 10^{-12} \gev & 4.8 \times 10^{-24} \gev & 2.4 \times 10^{-12} \gev \\
\hline m_\sigma & 30.2 \tev & 2.4 \times 10^8 \gev & 30.2 \tev & 2.4 \times 10^8 \gev \\
\hline g_\ast & 10.75 & 228.75 \; (\mbox{MSSM}) & 10.75 & 228.75 \; (\mbox{MSSM}) \\
\hline n & 1.4  & 1.4  & 1.1 & 1.1 \\
\hline M_{\mbox{\tiny max}} \;\; & 3.3 \times 10^{33} \; \mbox{g} & 7.5 \times 10^{22} \; \mbox{g}& 4.1 \times 10^{31}\; \mbox{g} & 1.6 \times 10^{20} \;\mbox{g} \\\hline M_{\mbox{\tiny min}} & 1.0 \times 10^{9} \; \mbox{g} & 1.3 \times 10^{5} \; \mbox{g} & 1.0 \times 10^{9} \; \mbox{g} & 1.3 \times 10^{5} \; \mbox{g} 
\\
\hline
\end{array} \nonumber 
\ee
\end{table*}

This conclusion is altered in a matter dominated phase.  Indeed, we saw in Section \ref{sectiongrow} that if the sound speed for the fluctuations vanishes (corresponding to no Jean's pressure) that the scalar perturbations grow.  In this case \eqref{rdubeta} no longer applies and PBH formation can occur significantly below the Hubble radius \cite{Carr:1975qj,Khlopov:1985jw,Polnarev:1986bi}.  
Modes that enter following the onset of scalar oscillations $H_{\mbox{\tiny osc}} \sim m_\sigma$ will grow linearly $\delta_\sigma \sim a(t)$ as discussed in Section \ref{sectiongrow}.  Given that the scalar oscillations evolve as pressureless matter (or `dust') the resulting mass fraction in PBHs is 
then \cite{Polnarev:1986bi}
\be \label{mdubeta}
\beta(M) \simeq 2 \times 10^{-2} \, \delta_M^{13/2}.
\ee
The suppression and scaling are determined by calculating the probability for an initial perturbation to collapse in a strictly spherically symmetric manner \cite{doroshkevich1970spatial}. Whereas non-spherical collapse would result in angular moment preventing further collapse below the Schwarzschild radius.  This expression gives the mass fraction during the non-thermal epoch, which begins near $t_{\mbox{\tiny osc}}\sim H^{-1} \simeq m^{-1}_\sigma$ and continues until the time of reheating $t_r\sim H^{-1} \sim \Gamma^{-1}$ when the scalar has mostly decayed to radiation and DM.  Because PBHs can form continuously during the non-thermal epoch there will be a range of masses with the maximal mass $M_{\mbox{\tiny max}}$ corresponding to the last PHB to form before the peak of scalar decay and reheating.
 As discussed in \cite{Polnarev:1986bi} it is convenient to express \eqref{mdubeta} in terms of the mass fraction of PBHs in a universe without a 
 non-thermal phase $\beta_0(M)$, since it is this quantity that is typically constrained,
 the two are related by \cite{Polnarev:1986bi}
\be \label{connectbeta}
 \beta(M)=\left( \frac{m_p}{\Gamma} \right)^{1/2} \left( \frac{m_p}{M} \right)^{1/2} \beta_0(M),
\ee
where the scaling accounts for the duration $t_{\mbox{\tiny osc}}<t<t_r$ in which the universe is matter dominated.
Given existing constraints on $\beta_0$  \cite{Carr:2009jm}  -- such as PBH evaporation leading to distortions in gamma-ray and the CMB and requiring relic PBHs to not exceed the critical 
density, we can place restrictions on the duration of the non-thermal phase using  \eqref{mdubeta} and \eqref{connectbeta}.
Combining these equations we find
\be \label{deltaM}
\delta_M = 1.8 \left( \frac{m_p^2}{\Gamma M} \right)^{1/13} \beta_0^{2/13}.
\ee
In the next section we will use constraints on $\beta_0$ to restrict model building in non-thermal histories.

\subsection{Expected Mass Range of PBHs}
The result of the last section \eqref{deltaM} provides an implicit constraint on model building from bounds on the mass fraction of PBHs $\beta_0$.
In practice, we need to establish the mass dependence of $\delta_M$, its relation to the primordial power spectrum and the 
range of expected PBH masses. 

Given the primordial power spectrum $P_\zeta \sim k^{n-1}$ we can relate this to
the density contrast through the Poisson equation on sub-Hubble scales implying $P_\delta \sim k^4 P_\zeta \sim k^{n+3}$.
We then use $\delta_k \sim P_\delta^{1/2}$, and that \eqref{Mdef} implies $M \sim k^{-3}$ for a matter dominated universe and we have
$\delta_M(t) \sim M^{-(n+3)/6}$.
Finally, recall that for usual PBH constraints, like those coming from \eqref{rdubeta}, the PBHs are taken to form on a particular scale ($M$) at the time of Hubble radius crossing.  Thus, we need to relate $\delta_M$ to its value at Hubble crossing.  Given the result of Section IIB that $\delta_k(t) \sim \delta_t(t_0) a(t)$ and using $k_H=a_H H \sim t^{-1/3}$ during matter domination we find
\be
\delta_M(t_{\mbox{\tiny H}}) \sim M^{(1-n)/6}.
\ee

The maximum mass allowed for a PBH during the non-thermal epoch will result from formation at the last moments before reheating $t_r$ when the Hubble horizon is largest.
However, unlike the standard PBH calculation we must account for the sub-horizon growth.
Thus, the largest mass PBH will correspond to $\delta_M(t_r)\sim {\cal O}(1)  $ where 
{\small \bea
\delta_M(t_r)&=&\delta_M(t_H)\left(\frac{a(t_r)}{a(t_H)}\right) =\delta_M(t_H)\left(\frac{t_r}{t_H}\right)^{2/3} \nonumber \\
&=&\delta_{\mbox{\tiny C}}\left(\frac{M_{\mbox{\tiny max}}}{M_{\mbox{\tiny C}}}\right)^{\frac{1-n}{6}}\left(\frac{t_r}{t_H}\right)^{2/3}
\sim {\cal O}(1), \;\;\;\;\;\;\;\;\; \label{res1}
\eea 
{\normalsize and}} we have introduced the COBE normalization $\delta_{\mbox{\tiny C}}\simeq 3.8\times10^{-6}$ (with $M_{\mbox{\tiny C}}$ the corresponding mass scale).  Evaluating \eqref{Mdef} at the time of horizon crossing in a matter dominated universe ($k_H=a_HH\sim t_H^{-1/3}$) 
and using this result for $t_H$ in \eqref{res1} we can solve for the maximum PHB mass
\bea
M_{\mbox{\tiny max}} 
&=& \alpha^{\frac{1}{n+3}}  \left(\frac{M_{\mbox{\tiny C}}}{m_p}\right)^{\frac{n-1}{n+3}}\left(\frac{m_{p}}{m_\sigma}\right)^{\frac{12}{n+3}} \, m_{p},  
\label{eq:mmax} 
\eea
where $\alpha=3.6 \times 10^{-22}$, 
$M_C=10^{57} h^{-1} $ g ($1$ \gev = $1.8 \times 10^{-24}$ g) with $h$ the Hubble parameter in units of $100$ km/s/Mpc (we take $h=0.7$), and we used the decay rate $\Gamma = m_\sigma^3 / m_p^2$.

The minimal PBH mass corresponds to the collapse of the Hubble volume at the onset of scalar oscillations (matter domination) $H_{\mbox{\tiny osc}} \simeq m_\sigma$.  
Using the Hubble equation and the energy density in the volume at that moment is $\rho =M_{\mbox{\tiny min}} H_{\mbox{\tiny osc}}^3$, we have
\be
M_{\mbox{\tiny min}} = \frac{3H_{\mbox{\tiny osc}}^2 m_p^2 }{H_{\mbox{\tiny osc}}^3}=3 \frac{m_p^2}{H_{\mbox{\tiny osc}}} = 3 \frac{m_p^2}{m_\sigma}.
\label{Mmin}
\ee
Using this result along with \eqref{eq:mmax} the allowed range of PBH formation falls within the range $M_{\mbox{\tiny min}}  \lesssim M \lesssim M_{\mbox{\tiny max}}$.  We note that both \eqref{eq:mmax} and \eqref{Mmin} are only a function of the scalar (moduli) mass.  In non-thermal SUSY WIMP models \cite{Acharya:2008bk,Acharya:2009zt,Kane:2015jia} this mass results from gravitational or anomaly mediated SUSY breaking and is proportional to the gravitino mass up to a factor of ${\cal O}(0.1-100)$ (which depends on the underlying parameters of the UV theory and is calculable). 
That is, $m_\sigma \sim m_{3/2}$ is a prediction of the theory and when SUSY breaking (at least partially) addresses the electroweak hierarchy problem we must require $m_{3/2} \sim$ TeV and so we expect $m_\sigma \sim 1 - 100$ TeV.  In models of Split SUSY \cite{ArkaniHamed:2004fb}, this requirement can be dropped since scalar masses can be far above the electroweak scale, but this implies the hierarchy problem must be addressed by other means.  In this paper we will consider both proposals, and we choose two fiducial models given in Table \ref{table1}.

The reheat temperature resulting from scalar decay in these models is $T_r \sim (\Gamma m_p)^{1/2}$.  As we review in Section \ref{secDM}, this temperature is important for establishing the expected microscopic properties of DM. We again see from the expression for the reheat temperature that the only new input into these models is the scalar mass $m_\sigma$, which makes a prediction for the expected range of PBH masses.

\subsection{Observational Constraints}
Given the allowed range of PBH masses we now compare to existing observational constraints.
There are a number of observations that may be used to constrain the mass fraction of PBHs -- for a review see \cite{Carr:2009jm}.  PBHs with masses around $10^{15}$ g will be evaporating today and lead to gamma-ray and cosmic-ray signals detectable by experiments like FERMI \cite{Ackermann:2015zua}. 
The evaporation time for a PHB is \cite{Carr:2009jm}
\be
t_{\mbox{\tiny evap}} \simeq 6.7 \times 10^{17} \;  \left(\frac{M_{\mbox{\tiny PBH}}}{10^{15} \; \mbox{g} }\right)^3 \; \mbox{s}.
\ee  
Thus, PBHs with masses far below $10^{15}$ g would have evaporated in the early universe and altered processes such as BBN.  Finally, PHBs with mass larger than $10^{15}$ g (as well as any remnants of the evaporated, light PBHs) are DM candidates and requiring their abundance to be in agreement with observations provides a constraint.  The authors of \cite{Carr:2009jm} have gathered all the existing bounds on PBHs for various mass ranges and we will use their data to constrain our models.   For the mass range of PBHs we are interested in here, the strongest bounds will result from BBN, CMB, and gamma-ray detection experiments.

Given the range of expected masses in Table \ref{table1}, we
see that a number of the PBHs will evaporate prior to BNN.  However, the strongest constraints come from the mass region near $10^{15}$ g -- PBHs evaporating today -- and these 
are restricted by gamma-ray observations.
Our results agrees with the analysis of \cite{Carr:1994ar}, except in that paper the authors took the lowest possible reheating temperature as the maximum 
temperature we are considering here. This is because the authors were concerned about the process of baryogenesis occurring, however this issue has been addresses in non-thermal WIMP models \cite{Kane:2011ih,Kane:2015jia} and so we take the lower bound on the reheat temperature to be set by BBN ($T \simeq 3 \mev$). For the low reheat model our constraints correspond to the far right side of Figure 5 in \cite{Carr:1994ar}, where gamma-ray constraints were the strongest. 

\begin{figure*}[t!]
\begin{center}
\includegraphics[scale=0.70]{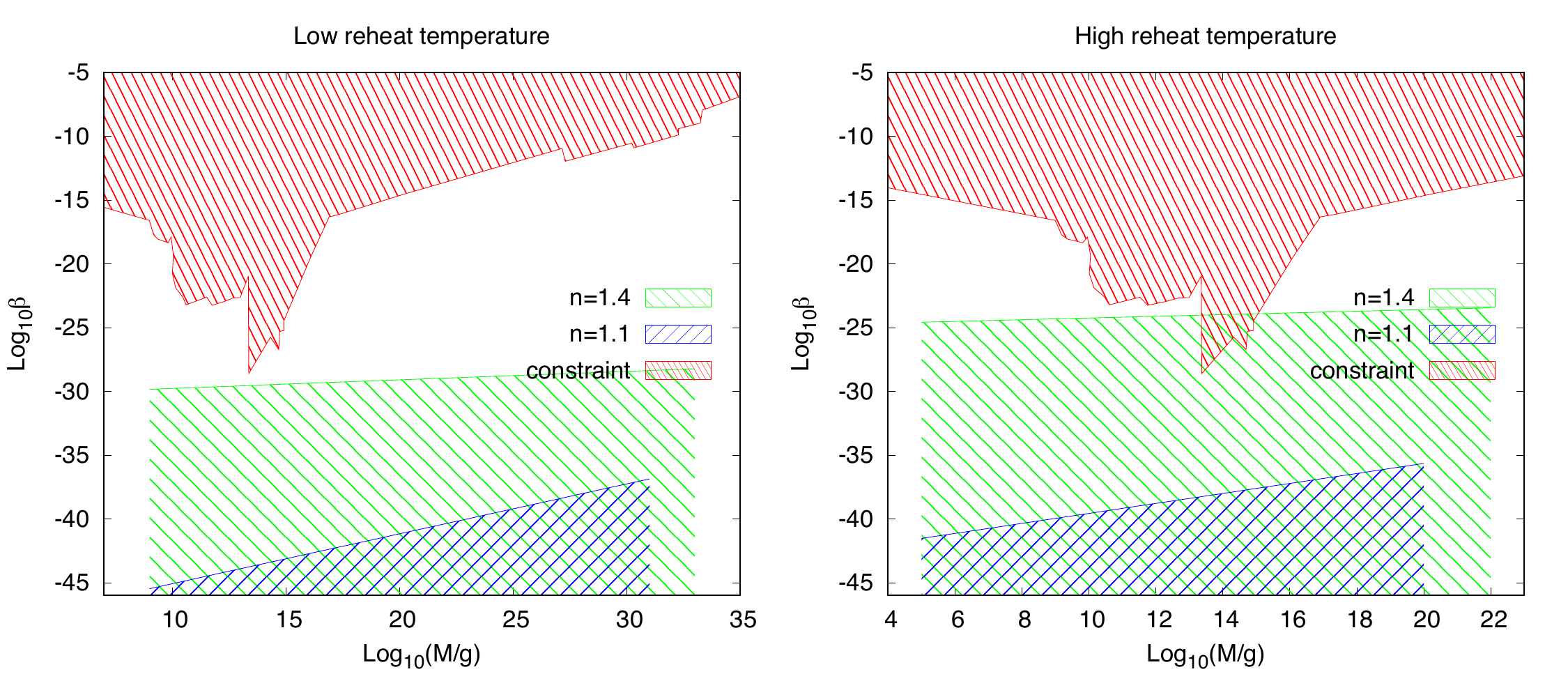}
\end{center}
\caption{\label{fig2} The left (right) plot gives the constraints for the low (high) reheat model discussed in the text (see also Table \ref{table1}). The restricted region is at the top (red) and the light (green) region represents models with a spectral tilt of $n=1.4$, whereas the darker lower region (blue) is nearly scale invariant at $n=1.1$.  We see that the stronger constraints are for high reheat models (right panel) where the matter dominated phase is shorter.  This is because the evolution in a matter dominated universe softens PBH constraints as the density of black holes and/or evaporation products will scale more slowly compared to that in a radiation dominated universe.  This effect is more important than the fact that PBHs have more time to form during a longer matter phase. The observational constraints are comprised of a number of different observations and taken from the review \cite{Carr:2009jm}.}
\end{figure*}

We now compare the predicted mass fraction of PBHs in our model to the constraints as summarized\footnote{We note that the constraints in \cite{Carr:2009jm} are related to our value of the mass fraction by 
$\beta'(M)\equiv \gamma^{1/2}\left(\frac{g_*}{106.75}\right)^{-1/4}\beta_0(M)$ where 
$\gamma$ is an order one number that depends on the details of the gravitational collapse, 
and our values for $g_*$ are given in Table~\ref{table1}.} in Fig. 9 of \cite{Carr:2009jm}. We calculate the mass fraction $\beta(M)$ in our model with \eqref{mdubeta} for two different values of the spectral index $n$. We follow the treatment in \cite{Carr:1994ar}, by normalizing $\delta_M$ in \eqref{mdubeta} to the CMB quadrupole so that $\delta_M = \delta_{\mbox{\tiny C}}  \, (M/M_{\mbox{\tiny C}})^{(1-n)/6}$.  As discussed above, in order to compare our results with the constraints given for a universe without a non-thermal phase, we have to relate the predicted ranges for $\beta$ to $\beta_0$ using \eqref{connectbeta}. 

Fig.~\ref{fig2} gives our results for the two different reheat temperatures and values of the spectral index. We see for a blue spectrum of $n=1.4$ we are overproducing PBHs in a mass range between $10^{13}$ g to $10^{15}$ g, with the strongest constraints resulting from PBHs which complete their evaporation process in the present epoch, i.e. have a mass of $\sim 10^{15}$ g and so would contribute to the extragalactic photon background as measured by  Fermi LAT \cite{Ackermann:2015zua}. 
From the FERMI data we can infer a limit on the density ratio $\Omega_{\mbox{\tiny PBH}}\leq(9.8\pm2.5)\times 10^{-9}$, which corresponds to a constraint $\beta(M_*)< 6\times 10^{-26}$ where $M_\ast \simeq 10^{15}$ g is the mass of PBH with an evaporation time of order the age of the universe. The same method can be applied to generate a constraint from galactic gamma-ray emissions. If PBHs of mass $\sim M_*$ are present in the galaxy we expect their density to be greater near the galactic core. Therefore, they should contribute to the galactic gamma-ray background and this contribution would be separable from the extragalactic background because of anisotropy. Analysis of the data suggests limits comparable to those of the above mentioned extragalactic background \cite{Carr:2009jm}. In Fig.~\ref{fig2}, the left side of the peak (mass range $2.5\times 10^{13}$ g $\leq M \leq 2.4\times 10^{14}$ g) corresponds to the constraints on PBHs which evaporate after recombination and dampen small-scale CMB anisotropies. This constraint has been calculated to be $\beta<3\times 10^{-30}\left({M} / {10^{13}\;\mathrm{ g}}\right ) ^{3.1}$ (see \cite{Carr:2009jm} for details).

As expected, the results shown in Fig.~\ref{fig2} imply that the bluer the spectrum the stronger the constraints.  This is because perturbations with a larger initial amplitude will more easily collapse to form PBHs after horizon entry. One may have anticipated that the low reheat scenario would have been more strongly constrained given the longer duration of the matter domination. However, this is not the case.
Instead the phase helps soften constraints compared to PBH production in a thermal universe because the change in the background evolution from a radiation dominated to a matter dominated universe changes the way in which the radiation and PBH densities scale. That is, for a matter dominated universe the fraction of the energy density in PBHs scales as $\rho_{\mbox{\tiny PBH}}/\rho_m \sim constant$, whereas in a radiation dominated universe the fraction grows $\rho_{\mbox{\tiny PBH}}/\rho_r \sim a(t)$. This change in the scaling is captured by \eqref{connectbeta} and noting that the decay rate scales with the reheat temperature as $\Gamma \sim T_r^2 / m_p$ (recall $\beta_0$ is the quantity data constrains and so we want to find $\beta_0$ for $\beta$ given by \eqref{mdubeta}) and we see that lower reheat temperatures result in lower values of the final mass fraction.  
In the next section we examine how our PBH constraints can be used to restrict non-thermal DM model building. 

\begin{figure}[t!] 
\begin{center}
\includegraphics[scale=0.63]{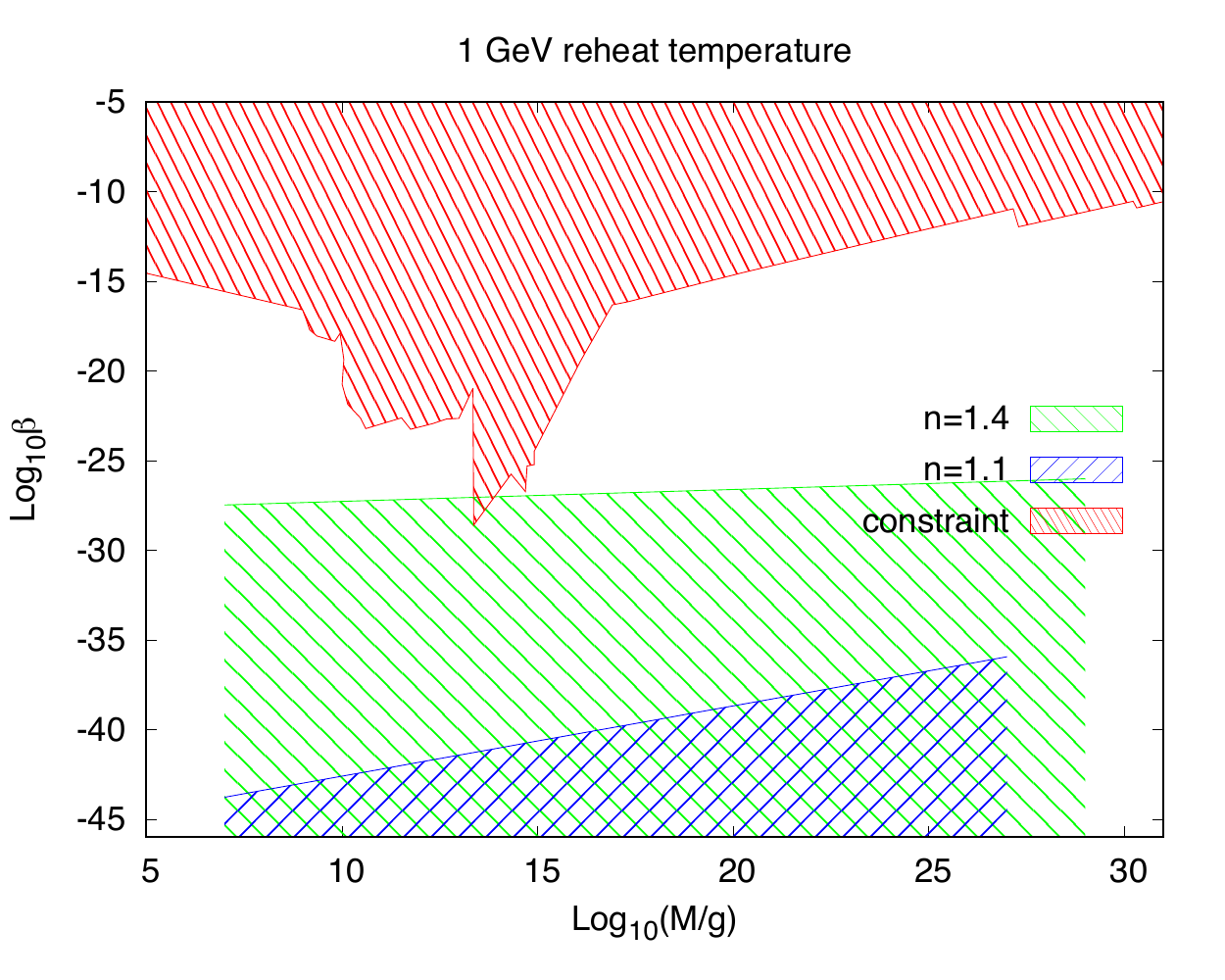}
\end{center}
\caption{\label{fig3} This plot gives the constraints relevant for the wino DM scenario discussed in the text. The restricted region is at the top (red) and the light (green) region represents models with a spectral tilt of $n=1.4$, whereas the darker lower region (blue) is nearly scale invariant at $n=1.1$. 
We have considered a GeV reheat temperature with $m_\sigma=1.4 \times 10^3$ TeV and $\Gamma=4.5 \times 10^{-19}$ GeV.  The allowed PBH mass range then has a minimal mass of $M_{\mbox{\tiny min}}=2.2 \times 10^7$ g, whereas the maximal masses are given by  $M_{\mbox{\tiny max}}=5.7 \times 10^{26}$ g and $M_{\mbox{\tiny max}}=6.8 \times 10^{28}$ g for $n=1.1$ and $n=1.4$, respectively.  We see that the formation of PBHs with masses around $M \simeq 10^{14}$ g are in tension with the data.}.
\end{figure}

\section{Implications for SUSY WIMPs \label{secDM}}
When the scalar decays and reheats the universe it dilutes any previous population of DM by a factor $\Omega_{\mbox{\tiny cdm}} \rightarrow \Omega_{\mbox{\tiny cdm}} (T_r/T_f)^3$, where $T_f$ is the thermal freeze-out temperature. This dilution is typically large enough to render the existing abundance irrelevant. This scaling can be understood from the fact that the temperature scales as\footnote{If there is significant decay of the scalar during the matter dominated epoch entropy is not conserved and this alters the temperature / scale factor relation \cite{Giudice:2000ex}. Here we will work in the instantaneous decay approximation and refer the reader to \cite{Fan:2014zua} for a more rigorous treatment. We do not expect any significant qualitative differences.} $T \sim 1/a(t)$, while volumes scale as $\sim a(t)^3$ (where $a(t)$ is the scale factor).  Although thermal DM will be diluted by scalar decay we also expect the scalar decay to produce DM. 

If the production of WIMPs leads to a number density that exceeds the critical value $n^c_x ={H}/{ \langle \sigma_xv \rangle} $ (evaluated at $T=T_r$), then the WIMPs will quickly annihilate down to this value, which acts as an attractor \cite{Acharya:2008bk}. The fixed point value is evaluated at the reheating temperature instead of the freeze-out temperature. This results in a parametric enhancement of the relic density 
\be \label{TandNTrelation}
\Omega^{NT}_{\mbox{\tiny cdm}} \, = \, \Omega^{T}_{\mbox{\tiny cdm}} \left( \frac{T_f}{T_r} \right) \,\,.
\ee
Since the relic density scales inversely with the annihilation cross section, this implies that WIMP candidates with annihilation cross sections larger than the canonical value by a factor $(T_f/T_r)$ can give the correct relic density in a non-thermal history with the above compensating factor. Examples of such WIMPs are the winos and higgsinos of the Minimal Supersymmetric extension of the Standard Model (MSSM). 

In more detail, the standard expression for the thermal relic density given by
\be  \label{thermal1}
\Omega_{\mbox{\tiny cdm}}^{T}h^2 \, = \, \frac{45}{2 \pi \sqrt{10}} \left(\frac{s_0}{\rho_c m_p}\right) \left(
\frac{m_X }{g^{1/2}_* \langle \sigma v \rangle T_f} \right),
\ee
where $\langle \sigma v \rangle$ is the DM self annihilation rate and $m_X$ is the DM mass.
Using (\ref{TandNTrelation}) and (\ref{thermal1}) we can estimate the relic density in non-thermal DM as
\bea  \label{nt1}
\Omega_{\mbox{\tiny cdm}}^{NT}h^2 &\simeq& 0.76 \, \left(\frac{s_0}{\rho_c m_p}\right) \left(
\frac{m_X }{g^{1/2}_* \langle \sigma v \rangle m_p T_r} \right) \,  \nonumber \\
\nn &\simeq& 0.10 \, \left( \frac{m_X}{100 \; \mbox{\small GeV}}\right) \left(\frac{10.75}{g_\ast} \right)^{1/4} \\
&\times&
 \left( \frac{3 \times 10^{-24} \; \mbox{cm}^3/\mbox{s}}{\langle \sigma v \rangle} \right) \left( \frac{\mbox{100 \,\mbox{\small TeV}}}{m_\sigma}\right)^{3/2}, \;\;\;\;\;\;
\eea
where the entropy density today is $s_0 = 2.78 \times 10^8 \rho_c/h^2$ and $h$ is again the Hubble parameter in units of $100$ km/s/Mpc.  We chose a fiducial value of $g_{*} = 10.75$ for the number of relativistic degrees of freedom at the time of reheating, although this would increase for reheat temperatures significantly above an MeV.

Unlike the standard thermal result, \eqref{nt1} depends on both the properties of the DM (mass and annihilation rate) and on the mass of the decaying scalar ($m_\sigma$).
 As discussed in the previous section the scalar mass (and so reheat temperature) is {\it not} a free parameter, but related to the gravitino mass ($m_{3/2}$).
The mass of the scalar (and so the relic density of DM) is controlled by the need for SUSY to generate a hierarchy between the electroweak and Planck scale.

We have chosen a fiducial value for the annihilation rate in \eqref{nt1} that yields roughly the right amount of DM  for the  hierarchy set by the choice of low-scale SUSY breaking.  The  cross-section is three orders of magnitude higher than expected with a thermal history with important experimental consequences, particularly for indirect detection of DM experiments.  Moreover, for SUSY WIMPs this suggests the wino-like neutralino is the preferred candidate, whereas in thermal scenarios the wino would not lead to the appropriate relic abundance.

Given the enhanced self-annihilation rate of the wino we can use existing constraints from indirect detection experiments (such as FERMI) to place constraints on the annihilation rate \cite{Easther:2013nga,Cohen:2013ama,Fan:2014zua}}.  In fact, if we consider single component DM, we can then use \eqref{nt1} to place restrictions on the reheat temperature (and so the scalar mass).  That is, \eqref{nt1} must agree with cosmological bounds set by Planck $\Omega_{\mbox{\tiny cdm}}h^2=0.1199 \pm 0.0027$ \cite{Ade:2015xua}, and then we can use FERMI constraints on the annihilation rate to bound the smallest allowed reheat temperature.  This was done in \cite{Easther:2013nga,Fan:2014zua}.  The authors of   \cite{Fan:2014zua} found the most rigid constraints by also including HESS data, even when carefully accounting for astrophysical uncertainties.  They found that for the wino to be a successful candidate the reheat temperature must exceed a GeV.  Given this result, we can now consider a GeV scale model and ask what PBH constraints imply.  

Fig. 3 gives the constraints relevant for the wino non-thermal DM scenario. The restricted region is at the top (red) and the light (green) region represents models with a spectral tilt of $n=1.4$, whereas the darker lower region (blue) is nearly scale invariant at $n=1.1$. 
We have considered a GeV reheat temperature with $m_\sigma=1.4 \times 10^3$ TeV and $\Gamma=4.5 \times 10^{-19}$ GeV.  The allowed PBH mass range corresponds to a minimal mass of 
$M_{\mbox{\tiny min}}=2.2 \times 10^7$ g, whereas the maximal masses are given by  $M_{\mbox{\tiny max}}=5.7 \times 10^{26}$ g and $M_{\mbox{\tiny max}}=6.8 \times 10^{28}$ g for $n=1.1$ and $n=1.4$, respectively.  We see that the formation of PBHs with masses around $M \simeq 10^{14}$ g are in tension with the data.  

Thus far we have considered constraints for fixed values of the reheating temperature and spectral index.  Now we consider constraints on the spectral index in SUSY WIMP models by combining our PBH constraints with existing constraints on DM.  Our results appear in Fig. 4 for a range of reheating temperatures (scalar masses $m_\sigma$).  
Regions with reheat temperatures below $1$ GeV are ruled out by both BBN constraints, as well as the FERMI/HESS constraints of \cite{Fan:2014zua}.  Temperatures above around $1000$ TeV are forbidden for non-thermal DM, since the mass must be less than this scale due to unitarity constraints and so DM is necessarily thermal in this region with $m_X \ll T_r$.
Thus, our main conclusion of this section is that restricting to models with non-thermal DM we find that PBH constraints disfavor an extremely blue spectrum and we find $n\lesssim 1.3 - 1.2$ for the spectral tilt.

\begin{figure}[t!] 
\begin{center}
\includegraphics[scale=0.63]{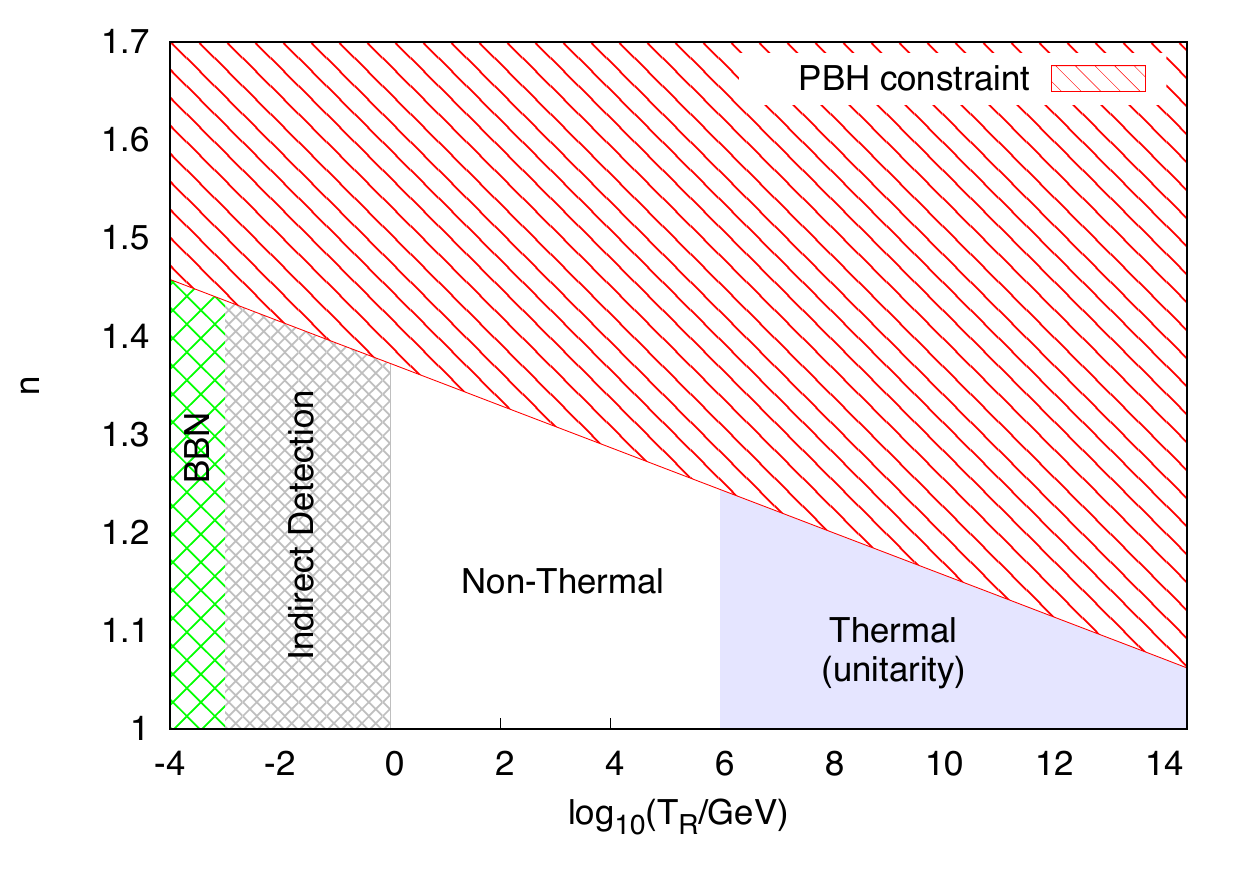}
\end{center}
\caption{\label{fig4} 
In this Figure we combine our PBH constraints with existing indirect detection bounds on non-thermal SUSY WIMP models.
Reheat temperatures below around $1$ GeV are ruled out by the indirect detection bounds given in \cite{Fan:2014zua}.  Above around $1000$ TeV the WIMPs would necessarily be thermal
since masses in excess of that scale would violate unitarity constraints and so in that region $T>m_X$ -- DM is thermal.  The top red region represents the PBH constraints from a variety
of observations including gamma-rays, CMB, and relic constraints all taken from the review \cite{Carr:2009jm}. }
\end{figure}

\section{Breakdown of Scalar Coherence? \label{breakdown}}
It is well known that a scalar coherently oscillating about the minimum of its potential scales as pressureless matter when averaged over a Hubble time \cite{Turner:1983he}.  
However, it is also known from both investigations into axions \cite{Marsh:2015xka} and in the context of scalar field DM / dark energy \cite{Hu:2000ke} that this approximation can fail on sub-horizon scales.  When this breakdown occurs we can still use the perturbation equations \eqref{fluid_eoms}, but the assumption of $w=c_s=0$ is no longer valid.
 In particular, \eqref{scalar1} and \eqref{scalar2} are no longer appropriate, and if a finite sound speed exists for the perturbations this implies a finite Jean's length $\lambda_J \sim c_s H^{-1}$, which can prevent the formation of structure on sub-Hubble scales -- including black holes.  
 
To consider the case of non-vanishing sound speed we begin by noting that for most of the non-thermal history the energy density is dominated by the scalar field as seen in Figure \ref{fig1}.  
Although the scalar density does not decrease noticeably during this epoch, it does not mean that decay and annihilation effects are not important (e.g. this changes the scale factor / temperature relation and how radiation and DM evolve), but for the growth of scalar perturbations these terms have little effect as long as the scalar is dominant \cite{Fan:2014zua} -- so in what follows we will neglect these terms.   A further simplification can be made if we take the sub-Hubble limit of these equations so that \eqref{EE1} reduces to the usual Poisson equation
\be \label{blah}
k^2 \Phi = - \frac{3}{2} H^2 a^2  \delta,
\ee
where for simplicity we take $\delta \equiv \delta_\sigma$ in what follows and taking the same 
limits of \eqref{fluid_eoms}, differentiating and combining the two equations, and using \eqref{blah} we find
\be
\ddot{\delta}+2H\dot{\delta}+\left( \frac{k^2 c_{\mbox{\tiny eff}}^2}{a^2} -\frac{3}{2}H^2 \right)\delta=0,
\ee
where $c_{\mbox{\tiny eff}}$ is the effective sound speed given by 
\be \label{soundspeed}
c_{\mbox{\tiny eff}}^2\equiv \langle c_s^2 \rangle_t = \frac{k^2/(4m_\sigma^2a^2)}{1+k^2/(4m_\sigma^2a^2)},
\ee
where we have taken an average which is valid for $t > m_\sigma^{-1}$ (recall $m_\sigma^{-1} \lesssim H^{-1}$ during oscillations) and the same average implies that $\langle w \rangle_t =\langle \dot{w} \rangle_t =0$ just as in 
the case of axions\footnote{We note that the sound speed $c_s^2 = \delta p / \delta \rho$ should not be confused with the adiabatic sound speed $c_{ad}^2 = \dot{p} / \dot{\rho}=w -\dot{w}/(3H(1+w))$ which vanishes when we take the average.  In particular, we see on sub-Hubble scales that the sound speed depends on momentum (is not homogeneous).  The two are related by $c_s^2=c_{ad}^2 + w \Gamma / \delta$ where $\Gamma$ is the non-adiabatic pressure perturbation. We refer the reader to the review  \cite{Marsh:2015xka} for more details.} \cite{Marsh:2015xka}.

We have seen that the strongest constraints come from the PBHs with mass near $M\simeq 10^{15}$ g.
We now estimate whether the pressure effects of a non-zero sound speed could stop the formation of these PBHs.
First we will assume the PBHs formed from the immediate collapse of modes as they entered the horizon, so that there mass will 
be set by the horizon mass $M \sim M_H = 3 m_p^2 / H$.  This would be the situation for a universe with finite equation of state \cite{Carr:1994ar}
and we will justify making this estimate below.

Given this assumption, we can solve to get the Hubble scale at this time
$H=3 m_p^2 M_H^{-1}$ and the modes at horizon crossing are $k=aH=a \, 3  m_p^2 M^{-1}$.
Using this in \eqref{soundspeed} the sound speed is then given by
\be \label{soundspeed2}
c_{\mbox{\tiny eff}}^{-2} = 1+\left( \frac{2M m_\sigma}{3 m_p^2} \right)^2,
\ee
which for very light scalars would imply $c_{\mbox{\tiny eff}} \simeq 1$ and this could prevent collapse.  
However, in both the low and high reheat models given in Table 1, the scalar must be quite heavy.
Using a PBH mass of $M = 10^{15}$ g we find an effective sound speed of 
\bea
c_{\mbox{\tiny eff}}^2&=& 2.7 \times 10^{-13} \ll 1  \;\;\; \mbox{(Low Reheat)}, \\
c_{\mbox{\tiny eff}}^2&=&  4.2 \times 10^{-21} \ll 1 \;\;\; \mbox{(High Reheat)}.
\eea
That is, the sound speed is negligible and pressure effects will not tend to prevent PBH formation.

In arriving at this conclusion we have assumed the perturbations collapse to form PBH as soon as they enter the Hubble radius.
We have seen this need not be the case and in fact it is the sub-Hubble evolution that leads to formation in non-thermal histories.
But in this case, the momentum in \eqref{soundspeed} will be even smaller (as it redshifts for each particular $k$).
Thus, the estimate of the sounds speed (and so pressure effects) given above are actually much more conservative than what we expect in practice.
The fact that the sound speed is negligible except for very small masses is well known from axion studies and warm DM \cite{Gorbunov:2011zzc}.
In conclusion, we find for the range of scalar mass and reheat temperatures considered here that the pressure of the scalar field results in a 
negligible effect on the formation of PBHs.

\section{Conclusions}
We have seen that in non-thermal histories the coherent oscillations of a scalar field lead to an effectively matter dominated universe, and so once density perturbations enter the horizon they will grow linearly with the scale factor. This growth leads to the formation of PBHs on sub-Hubble scales, contrary to the thermal case where PBHs only form from collapse near horizon crossing.  

We constrained the PBH abundance and their decay products through a number of experiments and observations, including gamma-ray, cosmic-ray and CMB observations.
The importance of each type of constraint depends on the PBH mass range, which we 
have calculated for the scalar masses (and so reheat temperatures) relevant for non-thermal WIMPs \cite{Kane:2015jia}.  The strongest constraints correspond to $10^{15}$ g PBHs, which are evaporating today and are constrained by FERMI. The constraints imply that non-thermal DM models and a primordial power spectrum with a significantly blue tilt are in conflict with the data. 
In particular, we have seen that when non-thermal SUSY WIMPs make up all of the DM, the tilt of the power spectrum at post-inflationary scales (prior to BBN) must satisfy $n \lesssim 1.25$ for high reheat temperatures. We saw that this constraint becomes weaker at low reheat temperatures, since the longer the duration of the non-thermal phase the weaker the constraints become due to the scaling of the critical density in PBHs compared to the thermal case.  In the former case the ratio is fixed, whereas in the thermal case $\rho_{\mbox{\tiny PBH}} / \rho_r \sim a(t)$ grows.

We also considered whether taking into account the breakdown of the coherence of the scalar oscillations on sub-Hubble scales could lead to pressure effects preventing the formation of PBHs.
Given the large mass of the scalar -- which is fixed by the underlying SUSY breaking -- we find that pressure effects are negligible.  
Our main conclusion is that when PBH constraints are combined with existing constraints on non-thermal DM model building that a primordial spectrum closer to scale invariance is preferred to that of a blue tilt.
\\

\section*{Acknowledgements}
We would like to thank Adrienne Erickcek, JiJi Fan, Justin Khoury, Doddy Marsh, Ogan \"Ozsoy, and Kuver Sinha for useful conversations, and we would especially like to thank Gary Gibbons for raising the question of whether PBH constraints are important for non-thermal histories.  GS and SW were supported in part by NASA Astrophysics Theory Grant NNH12ZDA001N, and DOE grant DE-FG02-85ER40237. 

\begin{thebibliography}{29}
\expandafter\ifx\csname natexlab\endcsname\relax\def\natexlab#1{#1}\fi
\expandafter\ifx\csname bibnamefont\endcsname\relax
  \def\bibnamefont#1{#1}\fi
\expandafter\ifx\csname bibfnamefont\endcsname\relax
  \def\bibfnamefont#1{#1}\fi
\expandafter\ifx\csname citenamefont\endcsname\relax
  \def\citenamefont#1{#1}\fi
\expandafter\ifx\csname url\endcsname\relax
  \def\url#1{\texttt{#1}}\fi
\expandafter\ifx\csname urlprefix\endcsname\relax\def\urlprefix{URL }\fi
\providecommand{\bibinfo}[2]{#2}
\providecommand{\eprint}[2][]{\url{#2}}

\bibitem[{\citenamefont{Acharya et~al.}(2008)\citenamefont{Acharya, Kumar,
  Bobkov, Kane, Shao, and Watson}}]{Acharya:2008bk}
\bibinfo{author}{\bibfnamefont{B.}~\bibnamefont{Acharya}},
  \bibinfo{author}{\bibfnamefont{P.}~\bibnamefont{Kumar}},
  \bibinfo{author}{\bibfnamefont{K.}~\bibnamefont{Bobkov}},
  \bibinfo{author}{\bibfnamefont{G.}~\bibnamefont{Kane}},
  \bibinfo{author}{\bibfnamefont{J.}~\bibnamefont{Shao}}, \bibnamefont{and}
  \bibinfo{author}{\bibfnamefont{S.}~\bibnamefont{Watson}},
  \bibinfo{journal}{JHEP} \textbf{\bibinfo{volume}{0806}}, \bibinfo{pages}{064}
  (\bibinfo{year}{2008}), \eprint{0804.0863}.

\bibitem[{\citenamefont{Acharya et~al.}(2009)\citenamefont{Acharya, Kane,
  Watson, and Kumar}}]{Acharya:2009zt}
\bibinfo{author}{\bibfnamefont{B.~S.} \bibnamefont{Acharya}},
  \bibinfo{author}{\bibfnamefont{G.}~\bibnamefont{Kane}},
  \bibinfo{author}{\bibfnamefont{S.}~\bibnamefont{Watson}}, \bibnamefont{and}
  \bibinfo{author}{\bibfnamefont{P.}~\bibnamefont{Kumar}},
  \bibinfo{journal}{Phys.Rev.} \textbf{\bibinfo{volume}{D80}},
  \bibinfo{pages}{083529} (\bibinfo{year}{2009}), \eprint{0908.2430}.

\bibitem[{\citenamefont{Kane et~al.}(2015{\natexlab{a}})\citenamefont{Kane,
  Sinha, and Watson}}]{Kane:2015jia}
\bibinfo{author}{\bibfnamefont{G.}~\bibnamefont{Kane}},
  \bibinfo{author}{\bibfnamefont{K.}~\bibnamefont{Sinha}}, \bibnamefont{and}
  \bibinfo{author}{\bibfnamefont{S.}~\bibnamefont{Watson}},
  \bibinfo{journal}{Int. J. Mod. Phys.} \textbf{\bibinfo{volume}{D24}},
  \bibinfo{pages}{1530022} (\bibinfo{year}{2015}{\natexlab{a}}),
  \eprint{1502.07746}.

\bibitem[{\citenamefont{Erickcek and Sigurdson}(2011)}]{Erickcek:2011us}
\bibinfo{author}{\bibfnamefont{A.~L.} \bibnamefont{Erickcek}} \bibnamefont{and}
  \bibinfo{author}{\bibfnamefont{K.}~\bibnamefont{Sigurdson}},
  \bibinfo{journal}{Phys.Rev.} \textbf{\bibinfo{volume}{D84}},
  \bibinfo{pages}{083503} (\bibinfo{year}{2011}), \eprint{1106.0536}.

\bibitem[{\citenamefont{Fan et~al.}(2014)\citenamefont{Fan, Ozsoy, and
  Watson}}]{Fan:2014zua}
\bibinfo{author}{\bibfnamefont{J.}~\bibnamefont{Fan}},
  \bibinfo{author}{\bibfnamefont{O.}~\bibnamefont{Ozsoy}}, \bibnamefont{and}
  \bibinfo{author}{\bibfnamefont{S.}~\bibnamefont{Watson}},
  \bibinfo{journal}{Phys. Rev.} \textbf{\bibinfo{volume}{D90}},
  \bibinfo{pages}{043536} (\bibinfo{year}{2014}), \eprint{1405.7373}.

\bibitem[{\citenamefont{Erickcek}(2015)}]{Erickcek:2015jza}
\bibinfo{author}{\bibfnamefont{A.~L.} \bibnamefont{Erickcek}},
  \bibinfo{journal}{Phys. Rev.} \textbf{\bibinfo{volume}{D92}},
  \bibinfo{pages}{103505} (\bibinfo{year}{2015}), \eprint{1504.03335}.

\bibitem[{\citenamefont{Erickcek et~al.}(2015)\citenamefont{Erickcek, Sinha,
  and Watson}}]{Erickcek:2015bda}
\bibinfo{author}{\bibfnamefont{A.~L.} \bibnamefont{Erickcek}},
  \bibinfo{author}{\bibfnamefont{K.}~\bibnamefont{Sinha}}, \bibnamefont{and}
  \bibinfo{author}{\bibfnamefont{S.}~\bibnamefont{Watson}}
  (\bibinfo{year}{2015}), \eprint{1510.04291}.

\bibitem[{\citenamefont{Kane et~al.}(2015{\natexlab{b}})\citenamefont{Kane,
  Kumar, Nelson, and Zheng}}]{Kane:2015qea}
\bibinfo{author}{\bibfnamefont{G.~L.} \bibnamefont{Kane}},
  \bibinfo{author}{\bibfnamefont{P.}~\bibnamefont{Kumar}},
  \bibinfo{author}{\bibfnamefont{B.~D.} \bibnamefont{Nelson}},
  \bibnamefont{and} \bibinfo{author}{\bibfnamefont{B.}~\bibnamefont{Zheng}}
  (\bibinfo{year}{2015}{\natexlab{b}}), \eprint{1502.05406}.

\bibitem[{\citenamefont{Khlopov}(2010)}]{Khlopov:2008qy}
\bibinfo{author}{\bibfnamefont{M.~{\relax Yu}.} \bibnamefont{Khlopov}},
  \bibinfo{journal}{Res. Astron. Astrophys.} \textbf{\bibinfo{volume}{10}},
  \bibinfo{pages}{495} (\bibinfo{year}{2010}), \eprint{0801.0116}.

\bibitem[{\citenamefont{Carr et~al.}(2010)\citenamefont{Carr, Kohri, Sendouda,
  and Yokoyama}}]{Carr:2009jm}
\bibinfo{author}{\bibfnamefont{B.~J.} \bibnamefont{Carr}},
  \bibinfo{author}{\bibfnamefont{K.}~\bibnamefont{Kohri}},
  \bibinfo{author}{\bibfnamefont{Y.}~\bibnamefont{Sendouda}}, \bibnamefont{and}
  \bibinfo{author}{\bibfnamefont{J.}~\bibnamefont{Yokoyama}},
  \bibinfo{journal}{Phys. Rev.} \textbf{\bibinfo{volume}{D81}},
  \bibinfo{pages}{104019} (\bibinfo{year}{2010}), \eprint{0912.5297}.

\bibitem[{\citenamefont{Green}(2015)}]{Green:2014faa}
\bibinfo{author}{\bibfnamefont{A.~M.} \bibnamefont{Green}},
  \bibinfo{journal}{Fundam. Theor. Phys.} \textbf{\bibinfo{volume}{178}},
  \bibinfo{pages}{129} (\bibinfo{year}{2015}), \eprint{1403.1198}.

\bibitem[{\citenamefont{Ade et~al.}(2015)}]{Ade:2015lrj}
\bibinfo{author}{\bibfnamefont{P.~A.~R.} \bibnamefont{Ade}}
  \bibnamefont{et~al.} (\bibinfo{collaboration}{Planck})
  (\bibinfo{year}{2015}), \eprint{1502.02114}.

\bibitem[{\citenamefont{Carr et~al.}(1994)\citenamefont{Carr, Gilbert, and
  Lidsey}}]{Carr:1994ar}
\bibinfo{author}{\bibfnamefont{B.~J.} \bibnamefont{Carr}},
  \bibinfo{author}{\bibfnamefont{J.}~\bibnamefont{Gilbert}}, \bibnamefont{and}
  \bibinfo{author}{\bibfnamefont{J.~E.} \bibnamefont{Lidsey}},
  \bibinfo{journal}{Phys.Rev.} \textbf{\bibinfo{volume}{D50}},
  \bibinfo{pages}{4853} (\bibinfo{year}{1994}), \eprint{astro-ph/9405027}.

\bibitem[{\citenamefont{Khlopov et~al.}(1985)\citenamefont{Khlopov, Malomed,
  and Zeldovich}}]{Khlopov:1985jw}
\bibinfo{author}{\bibfnamefont{M.}~\bibnamefont{Khlopov}},
  \bibinfo{author}{\bibfnamefont{B.~A.} \bibnamefont{Malomed}},
  \bibnamefont{and} \bibinfo{author}{\bibfnamefont{I.~B.}
  \bibnamefont{Zeldovich}}, \bibinfo{journal}{Mon. Not. Roy. Astron. Soc.}
  \textbf{\bibinfo{volume}{215}}, \bibinfo{pages}{575} (\bibinfo{year}{1985}).

\bibitem[{\citenamefont{Polnarev and Khlopov}(1985)}]{Polnarev:1986bi}
\bibinfo{author}{\bibfnamefont{A.}~\bibnamefont{Polnarev}} \bibnamefont{and}
  \bibinfo{author}{\bibfnamefont{M.~Y.} \bibnamefont{Khlopov}},
  \bibinfo{journal}{Sov.Phys.Usp.} \textbf{\bibinfo{volume}{28}},
  \bibinfo{pages}{213} (\bibinfo{year}{1985}).

\bibitem[{\citenamefont{Giudice et~al.}(2001)\citenamefont{Giudice, Kolb, and
  Riotto}}]{Giudice:2000ex}
\bibinfo{author}{\bibfnamefont{G.~F.} \bibnamefont{Giudice}},
  \bibinfo{author}{\bibfnamefont{E.~W.} \bibnamefont{Kolb}}, \bibnamefont{and}
  \bibinfo{author}{\bibfnamefont{A.}~\bibnamefont{Riotto}},
  \bibinfo{journal}{Phys.Rev.} \textbf{\bibinfo{volume}{D64}},
  \bibinfo{pages}{023508} (\bibinfo{year}{2001}), \eprint{hep-ph/0005123}.

\bibitem[{\citenamefont{Mukhanov}(2005)}]{Mukhanov:2005sc}
\bibinfo{author}{\bibfnamefont{V.}~\bibnamefont{Mukhanov}},
  \emph{\bibinfo{title}{{Physical Foundations of Cosmology}}}
  (\bibinfo{publisher}{Cambridge University Press}, \bibinfo{address}{Oxford},
  \bibinfo{year}{2005}), ISBN \bibinfo{isbn}{0521563984, 9780521563987},
  \urlprefix\url{http://www-spires.fnal.gov/spires/find/books/www?cl=QB981.M89::2005}.

\bibitem[{\citenamefont{Young et~al.}(2014)\citenamefont{Young, Byrnes, and
  Sasaki}}]{Young:2014ana}
\bibinfo{author}{\bibfnamefont{S.}~\bibnamefont{Young}},
  \bibinfo{author}{\bibfnamefont{C.~T.} \bibnamefont{Byrnes}},
  \bibnamefont{and} \bibinfo{author}{\bibfnamefont{M.}~\bibnamefont{Sasaki}},
  \bibinfo{journal}{JCAP} \textbf{\bibinfo{volume}{1407}}, \bibinfo{pages}{045}
  (\bibinfo{year}{2014}), \eprint{1405.7023}.

\bibitem[{\citenamefont{Carr}(1975)}]{Carr:1975qj}
\bibinfo{author}{\bibfnamefont{B.~J.} \bibnamefont{Carr}},
  \bibinfo{journal}{Astrophys. J.} \textbf{\bibinfo{volume}{201}},
  \bibinfo{pages}{1} (\bibinfo{year}{1975}).

\bibitem[{\citenamefont{Doroshkevich}(1970)}]{doroshkevich1970spatial}
\bibinfo{author}{\bibfnamefont{A.}~\bibnamefont{Doroshkevich}},
  \bibinfo{journal}{Astrophysics} \textbf{\bibinfo{volume}{6}},
  \bibinfo{pages}{320} (\bibinfo{year}{1970}).

\bibitem[{\citenamefont{Arkani-Hamed and
  Dimopoulos}(2005)}]{ArkaniHamed:2004fb}
\bibinfo{author}{\bibfnamefont{N.}~\bibnamefont{Arkani-Hamed}}
  \bibnamefont{and}
  \bibinfo{author}{\bibfnamefont{S.}~\bibnamefont{Dimopoulos}},
  \bibinfo{journal}{JHEP} \textbf{\bibinfo{volume}{0506}}, \bibinfo{pages}{073}
  (\bibinfo{year}{2005}), \eprint{hep-th/0405159}.

\bibitem[{\citenamefont{Ackermann et~al.}(2015)}]{Ackermann:2015zua}
\bibinfo{author}{\bibfnamefont{M.}~\bibnamefont{Ackermann}}
  \bibnamefont{et~al.} (\bibinfo{collaboration}{Fermi-LAT}),
  \bibinfo{journal}{Phys. Rev. Lett.} \textbf{\bibinfo{volume}{115}},
  \bibinfo{pages}{231301} (\bibinfo{year}{2015}), \eprint{1503.02641}.

\bibitem[{\citenamefont{Kane et~al.}(2011)\citenamefont{Kane, Shao, Watson, and
  Yu}}]{Kane:2011ih}
\bibinfo{author}{\bibfnamefont{G.}~\bibnamefont{Kane}},
  \bibinfo{author}{\bibfnamefont{J.}~\bibnamefont{Shao}},
  \bibinfo{author}{\bibfnamefont{S.}~\bibnamefont{Watson}}, \bibnamefont{and}
  \bibinfo{author}{\bibfnamefont{H.-B.} \bibnamefont{Yu}},
  \bibinfo{journal}{JCAP} \textbf{\bibinfo{volume}{1111}}, \bibinfo{pages}{012}
  (\bibinfo{year}{2011}), \eprint{1108.5178}.

\bibitem[{\citenamefont{Easther et~al.}(2013)\citenamefont{Easther, Galvez,
  Ozsoy, and Watson}}]{Easther:2013nga}
\bibinfo{author}{\bibfnamefont{R.}~\bibnamefont{Easther}},
  \bibinfo{author}{\bibfnamefont{R.}~\bibnamefont{Galvez}},
  \bibinfo{author}{\bibfnamefont{O.}~\bibnamefont{Ozsoy}}, \bibnamefont{and}
  \bibinfo{author}{\bibfnamefont{S.}~\bibnamefont{Watson}}
  (\bibinfo{year}{2013}), \eprint{1307.2453}.

\bibitem[{\citenamefont{Cohen et~al.}(2013)\citenamefont{Cohen, Lisanti,
  Pierce, and Slatyer}}]{Cohen:2013ama}
\bibinfo{author}{\bibfnamefont{T.}~\bibnamefont{Cohen}},
  \bibinfo{author}{\bibfnamefont{M.}~\bibnamefont{Lisanti}},
  \bibinfo{author}{\bibfnamefont{A.}~\bibnamefont{Pierce}}, \bibnamefont{and}
  \bibinfo{author}{\bibfnamefont{T.~R.} \bibnamefont{Slatyer}},
  \bibinfo{journal}{JCAP} \textbf{\bibinfo{volume}{1310}}, \bibinfo{pages}{061}
  (\bibinfo{year}{2013}), \eprint{1307.4082}.

\bibitem{Ade:2015xua} 
  P.~A.~R.~Ade {\it et al.} [Planck Collaboration],
  arXiv:1502.01589 [astro-ph.CO].
  
\bibitem[{\citenamefont{Turner}(1983)}]{Turner:1983he}
\bibinfo{author}{\bibfnamefont{M.~S.} \bibnamefont{Turner}},
  \bibinfo{journal}{Phys.Rev.} \textbf{\bibinfo{volume}{D28}},
  \bibinfo{pages}{1243} (\bibinfo{year}{1983}).

\bibitem[{\citenamefont{Marsh}(2015)}]{Marsh:2015xka}
\bibinfo{author}{\bibfnamefont{D.~J.~E.} \bibnamefont{Marsh}}
  (\bibinfo{year}{2015}), \eprint{1510.07633}.

\bibitem[{\citenamefont{Hu et~al.}(2000)\citenamefont{Hu, Barkana, and
  Gruzinov}}]{Hu:2000ke}
\bibinfo{author}{\bibfnamefont{W.}~\bibnamefont{Hu}},
  \bibinfo{author}{\bibfnamefont{R.}~\bibnamefont{Barkana}}, \bibnamefont{and}
  \bibinfo{author}{\bibfnamefont{A.}~\bibnamefont{Gruzinov}},
  \bibinfo{journal}{Phys.Rev.Lett.} \textbf{\bibinfo{volume}{85}},
  \bibinfo{pages}{1158} (\bibinfo{year}{2000}), \eprint{astro-ph/0003365}.

\bibitem[{\citenamefont{Gorbunov and Rubakov}(2011)}]{Gorbunov:2011zzc}
\bibinfo{author}{\bibfnamefont{D.~S.} \bibnamefont{Gorbunov}} \bibnamefont{and}
  \bibinfo{author}{\bibfnamefont{V.~A.} \bibnamefont{Rubakov}},
  \emph{\bibinfo{title}{{Introduction to the theory of the early universe:
  Cosmological perturbations and inflationary theory}}} (\bibinfo{year}{2011}),
  \urlprefix\url{http://www.DESY.eblib.com/patron/FullRecord.aspx?p=737613}.

\end{thebibliography}

\end{document}